\newcommand{\beq}{\begin{equation}}
\newcommand{\eeq}{\end{equation}}
\newcommand{\beqa}{\begin{eqnarray}}
\newcommand{\eeqa}{\end{eqnarray}}
\newcommand{\ba}{\begin{array}}
\newcommand{\ea}{\end{array}}

\documentclass[pra,twocolumn,showpacs,preprintnumbers,
amsmath,amssymb,floatfix]{revtex4}
\usepackage{hyperref}
\usepackage{epsfig}

\begin{document}

\title{Condensate formation with three-component ultracold fermions} 

\author{Luca Salasnich}
\affiliation{INO-CNR, Dipartimento di Fisica ``Galileo Galilei'', 
Universit\`a di Padova, Via Marzolo 8, 35131 Padova, Italy}

\date{\today}

\begin{abstract}
We investigate the formation of Bose-Einstein condensation and population 
imbalance in a three-component Fermi superfluid by increasing 
the U(3) invariant attractive interaction. We consider 
the system at zero temperature in three dimensions and also 
in two dimensions. Within the mean-field theory, we derive 
explicit formulas for number densities, gap order parameter, 
condensate density and condensate fraction of the uniform system,   
and analyze them in the crossover 
from the Bardeen-Cooper-Schrieffer (BCS) state 
of Cooper pairs to the Bose-Einstein Condensate (BEC) 
of strongly-bound molecular dimers. In addition, we study this  
Fermi mixture trapped by a harmonic potential: we calculate 
the density profiles of the three components and the condensate density 
profile of Cooper pairs in the BCS-BEC crossover. 
\end{abstract}

\pacs{03.75.Hh, 03.75.Ss}

\maketitle

\section{Introduction} 

In the last years two experimental groups \cite{zwierlein,ueda}
have analyzed the condensate fraction of ultra-cold 
two hyperfine component Fermi vapors of $^6$Li atoms in 
the crossover from the Bardeen-Cooper-Schrieffer (BCS) 
state of Cooper Fermi pairs to the Bose-Einstein condensate (BEC) 
of molecular dimers. These experiments are in quite good agreement 
with mean-field theoretical predictions 
\cite{sala-odlro,ortiz} and Monte-Carlo 
simulations \cite{astrakharchik} at zero temperature, 
while at finite temperature beyond-mean-field 
corrections are needed \cite{ohashi}. 
\par 
Degenerate three-component gases have been 
experimentally realized using the three lowest hyperfine states 
of $^6$Li \cite{selim,hara}. At high magnetic fields 
the scattering lengths of this three-component system are 
very close each other and the system is approximately 
U(3) invariant. Moreover, it has been theoretically 
predicted that good $SU(N)$ invariance (with $N\leq 10$) 
can be reached with ultracold alkaline-earth 
atoms (e.g. with $^{87}$Sr atoms) \cite{ana}. Very recently, 
Ozawa and Baym have investigated the BCS-BEC crossover 
of uniform three-component ultracold fermions with U(3) symmetry 
at zero and finite temperature in three-dimensional space \cite{baym}. 
\par 
In this paper we focus on the condensate formation 
in a uniform three-component Fermi superfluid by increasing 
the U(3) invariant attractive interaction. We determine 
the condensate fraction and the population imbalance of the system 
in both the three-dimensional case and in the  
two-dimensional one. By using the extended BEC theory 
\cite{marini}, we obtain analytical formulas for 
number densities, condensate density 
and condensate fraction \cite{sala-odlro,sala-odlro2} 
of the uniform system and analyze them in the crossover 
from the Bardeen-Cooper-Schrieffer (BCS) state 
of Cooper pairs to the Bose-Einstein Condensate (BEC) 
of strongly-bound molecular dimers. We consider also 
the inclusion of a harmonic confinement, determining 
the density profiles of the three components of the ultracold gas, 
and the space dependence of condensate density and gap order 
parameter. In the paper we neglect effects of three-body interactions 
like Efimov three-body bound states. 

\section{The model} 

The shifted Hamiltonian density of a dilute and 
interacting three-component Fermi 
gas in a volume $V$ is given by \cite{baym}
\beqa 
{\hat {\cal H}}' &=& \sum_{\alpha=R,G,B} {\hat \psi}^+_{\alpha} 
\left(-{\hbar^2\over 2 m}\nabla^2- \mu \right) {\hat \psi}_{\alpha} 
+ g \big( {\hat \psi}^+_{R}{\hat \psi}^+_{G}
{\hat \psi}_{G}{\hat \psi}_{R} 
\nonumber
\\
&+&{\hat \psi}^+_{R}{\hat \psi}^+_{B}
{\hat \psi}_{B}{\hat \psi}_{R} 
+{\hat \psi}^+_{G}{\hat \psi}^+_{B}
{\hat \psi}_{B}{\hat \psi}_{G} \big) \; , 
\label{ham} 
\eeqa 
where ${\hat \psi}_{\alpha}({\bf r})$ is the field operator 
that destroys a fermion of component $\alpha$ 
in the position ${\bf r}$, while ${\hat \psi}_{\alpha}^+({\bf r})$ 
creates a fermion of component $\alpha$ in ${\bf r}$. 
To mimic QCD the three components are thought as three colors: red (R), 
green (G) and blue (B). 
The attractive inter-atomic interaction is described by a contact 
pseudo-potential of strength $g$ ($g<0$). 
The number density operator is 
\beq 
{\hat n}({\bf r}) = \sum_{\alpha=R,G,B} 
{\hat \psi}^+_{\alpha}({\bf r}){\hat \psi}_{\alpha}({\bf r}) 
\eeq  
and the average number of fermions reads 
\beq 
N=\int d^3{\bf r}\, \langle {\hat n}({\bf r})\rangle \; . 
\eeq
This total number $N$ is fixed by the chemical potential $\mu$ 
which appears in Eq. (\ref{ham}). As stressed in Ref. \cite{baym}, 
by fixing only the total chemical potential $\mu$ 
(or equivalently only the total number of atoms $N$) 
the Hamiltonian (\ref{ham}) is invariant under global U(3) rotations 
of the species (and consequently also under global SU(3)). 
At zero temperature, the attractive interaction leads to pairing 
of fermions which breaks the U(3) symmetry but only two colors 
are paired and one is left unpaired \cite{baym}. 
\par 
Following Ozawa and Baym \cite{baym}, we assume without 
loss of generality that the red and green particles are paired 
and the blue are not paired. The interacting terms can be then treated 
within the minimal mean-field BCS approximation, giving 
\beqa
g \ {\hat \psi}^+_{R}
{\hat \psi}^+_{G}
{\hat \psi}_{G}
{\hat \psi}_{R} 
= 
g \ \langle 
{\hat \psi}^+_{R}{\hat \psi}^+_{G}
\rangle 
{\hat \psi}_{G}
{\hat \psi}_{R} 
+ 
g \ {\hat \psi}^+_{R}
{\hat \psi}^+_{G} 
\langle 
{\hat \psi}_{G}{\hat \psi}_{R} 
\rangle 
\eeqa 
and 
\beq 
g \ {\hat \psi}^+_{R}
{\hat \psi}^+_{B}
{\hat \psi}_{B}
{\hat \psi}_{R} =  g \ {\hat \psi}^+_{G}
{\hat \psi}^+_{B} 
{\hat \psi}_{B}
{\hat \psi}_{G} = 0  \; . 
\eeq
Notice that the Hartree terms have been neglected, while 
the pairing gap $\Delta=g \langle 
{\hat \psi}_{G}{\hat \psi}_{R} 
\rangle$ between red and green fermions is the key quantity. 

The shifted Hamiltonian density (\ref{ham}) 
is diagonalized by using the Bogoliubov-Valatin  
representation of the field operator 
${\hat \psi}_{\alpha}({\bf r})$ 
in terms of the anticommuting quasi-particle 
Bogoliubov operators ${\hat b}_{{\bf k}\alpha}$ 
with amplitudes $u_k$ and $v_k$ and energy $E_k$. 
After minimization of the resulting quadratic Hamiltonian 
one finds familiar expressions for these quantities: 
\beq
E_k=\left[\left({\hbar^2k^2\over 2m}-\mu \right)^2 
+ \Delta^2\right]^{1/2}
\eeq
and
\beq \label{vk}
v_k^2 = {1\over 2} \left( 1 - \frac{{\hbar^2k^2\over 2m} 
- \mu }{E_k} \right) \, ,  
\eeq 
with $u_k^2=1 - v_k^2$. In addition we find 
the equation for the number of particles 
\beq 
N = N_R + N_G + N_B \; , 
\label{bcs1} 
\eeq 
where 
\beq 
N_R = N_G = {1\over 2} \sum_{\bf k} v_k^2 
\label{bcs1a}
\eeq
and 
\beq 
N_B = \sum_{\bf k} \Theta\left(\mu-{\hbar^2k^2\over 2m}\right) 
\label{bcs1b}
\; , 
\eeq
with $\Theta(x)$ the Heaviside step function, and also the gap equation 
\beq
-{1\over g} = {1 \over V} \sum_{\bf k} {1\over 2 E_k} \; . 
\label{bcsGap}
\eeq 
The chemical potential $\mu$ and the gap energy $\Delta$ are obtained by
solving equations (\ref{bcs1}) and (\ref{bcsGap}). 

We observe that the number of red-green pairs 
in the lowest state, i.e. the condensate 
number of red-green pairs, is given by \cite{sala-odlro,ortiz,ohashi}
\beq
N_0 = \int d^3{\bf r}_1 \; d^3{\bf r}_2 \; | \langle 
{\hat \psi}_{G}({\bf r}_1)  
{\hat \psi}_{R}({\bf r}_2)  
\rangle |^2 ,
\eeq 
and it is straightforward to show that 
\beq 
N_0 = \sum_{\bf k} u_k^2 v_k^2 \; . 
\eeq 

In the continuum limit, due to the 
choice of a contact potential, the gap equation (\ref{bcsGap}) 
diverges in the ultraviolet. This divergence is 
linear in three dimensions and logarithmic in two dimensions. 
Let us face this problem in the next two sections. 

\section{Three-component ultracold Fermions in three dimensions} 

In three dimensions, a suitable regularization \cite{marini} 
is obtained by introducing 
the inter-atomic scattering length $a_F$ via the equation 
\beq
-{1\over g} = - {m \over 4 \pi \hbar^2 a_F} + 
{1 \over V} \sum_{\bf k} \frac{m}{\hbar^2k^2} \,,
\eeq 
and then subtracting this equation from 
the gap equation (\ref{bcsGap}). 
In this way one obtains the three-dimensional regularized gap equation 
\beq 
-{m \over 4 \pi \hbar^2 a_F} = {1 \over V} 
\sum_{\bf k} \left( { 1 \over 2 E_k} 
- \frac{m}{\hbar^2k^2} 
\right) . 
\label{bcs3d}  
\eeq 
In the three-dimensional continuum limit 
$\sum_{\bf k} \to V/(2\pi)^3 \int d^3{\bf k}
\to V/(2\pi^2) \int k^2 dk$ 
from  the number equation (\ref{bcs1}) with (\ref{bcs1a}) 
and (\ref{bcs1b}) we find the total number density as 
\beq  
n = {N\over V} = n_R + n_G + n_B \; , 
\eeq 
with 
\beq 
n_R = n_G = {1\over 2} {(2m)^{3/2} \over 2 \pi^2 \hbar^3} \,
\Delta^{3/2} \, I_2\!\left({\mu \over \Delta}\right)  \, ,
\label{gbcs2} 
\eeq 
and 
\beq 
n_B = {1\over 3} 
{(2m)^{3/2} \over 2 \pi^2 \hbar^3} \mu^{3/2} \ \Theta(\mu) \; . 
\eeq
The renormalized gap equation (\ref{bcs3d}) becomes instead 
\beq 
-{1\over a_F} = {2 (2m)^{1/2} \over \pi \hbar^3} \,
\Delta^{1/2} \, I_1\!\left({\mu \over \Delta}\right)  \, , 
\label{gbcs1} 
\eeq 
where $k_F=(6\pi N/(3V))^{1/3}=(2\pi^2n)^{1/3}$ is the Fermi wave number.  
Here $I_1(x)$ and $I_2(x)$ are the two monotonic 
functions 
\beq 
I_1(x) = \int_0^{+\infty} y^2 
\left( {1\over \sqrt{(y^2-x)^2+1}} - {1\over y^2} \right) dy \; , 
\eeq
\beq 
I_2(x) = \int_0^{+\infty} y^2 \left( 1 - {y^2-x\over \sqrt{(y^2-x)^2+1}}
\right) dy \; , 
\eeq
which can be expressed in terms of elliptic 
integrals, as shown by Marini, Pistolesi 
and Strinati \cite{marini}. In a similar way we get 
the condensate density of the red-green pair as 
\beq 
n_0 = {N_0\over V} = {m^{3/2} \over 8 \pi \hbar^3} \,
\Delta^{3/2} \sqrt{{\mu\over \Delta}+\sqrt{1+{\mu^2 \over \Delta^2} }} 
\label{con0} \; .   
\eeq 
This equation and the gap equation (\ref{gbcs1}) are the same 
of the two-component 
superfluid fermi gas (see \cite{sala-odlro}) but the number 
equation (\ref{bcs1}), with (\ref{bcs1a}) and (\ref{bcs1b}), is 
clearly different. Note that all the relevant quantities can be expressed 
in terms of the ratio 
\beq 
x_0 = {\mu\over \Delta} \; . 
\eeq
For instance, the fraction of red fermions, which is equal to 
the fraction of green fermions, is given by 
\beq 
{n_R\over n} = {n_G\over n} = 
{I_2(x_0)\over 2 I_2(x_0) 
+ {2\over 3} x_0^{3/2}\ \Theta(x_0) } \; ,  
\eeq
while the fraction of blue fermions reads 
\beq 
{n_B\over n} = 1 - {I_2(x_0)\over I_2(x_0) 
+ {1\over 3} x_0^{3/2}\ \Theta(x_0) } \; . 
\eeq
The fraction of condensed red-green pairs is instead 
\beq 
{n_0\over n} = {\pi \over 8\sqrt{2}} 
{\sqrt{x_0+\sqrt{1+x_0^2}}\over I_2(x_0)+{1\over 3}
x_0^{3/2}\ \Theta(x_0) } \; . 
\eeq
Finally, the adimensional interaction strength of the BCS-BEC 
crossover is given by 
\beq 
y = {1\over k_F a_F} = -{2\over \pi} 
{I_1(x_0)\over  2 I_2(x_0) + {2\over 3} x_0^{3/2}\ \Theta(x_0) } \; . 
\eeq
We can use these parametric formulas of $x_0$ to plot the 
density fractions 
as a function of the scaled interaction strength $y$. 

\begin{figure}
\centerline{\epsfig{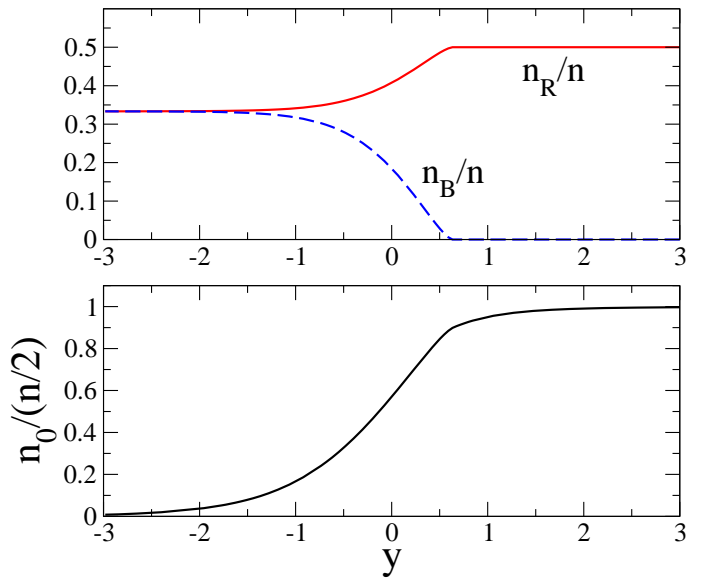}}
\small 
\caption{(Color online). Three-component ultracold fermions 
in three-dimensions. 
Upper panel: fraction of red fermions $n_R/n$ (solid line) 
and fraction of blue fermions $n_B/n$ (dashed line) 
as a function of scaled interaction strength $y=1/(k_Fa_F)$. 
Lower panel: condensed fraction of red-green particles $n_0/n$ 
as a function of scaled interaction strength $y=1/(k_Fa_F)$.} 
\label{fig1}
\end{figure} 

In the upper panel of Fig. \ref{fig1} we plot the fraction 
of red fermions $n_R/n$ (solid line) and the 
fraction of blue fermions $n_B/n$ (dashed line) 
as a function of scaled interaction strength $y=1/(k_Fa_F)$. 
The behavior of $n_G/n$ is not shown because it is exactly 
the same of $n_R/n$. The figure shows that 
in the deep BCS regime ($y\ll -1$) 
the system has $n_R/n=n_G/n=n_B/n=1/3$. By increasing $y$ 
the fraction of red and green fermions increases while 
the fraction of blue fermions decreases. At $y\simeq 0.6$, where $x_0=0$, 
the fraction of blue fermions becomes zero, i.e. $n_B/n=0$ 
and consequently $n_R/n=n_G/n=1/2$. For larger 
values of $y$ there are only the paired red and green particles. 
This behavior is fully consistent with the findings of 
Ozawa and Baym \cite{baym}. 

In the lower panel of Fig.\ref{fig1} 
it is shown the plot of the condensate fraction $n_0/(n/2)$ of 
red-green pairs through the BCS-BEC crossover as a function of 
the Fermi-gas parameter $y=1/(k_Fa_F)$. 
The figure shows that a large condensate fraction builds up in the 
BCS side already before the unitarity limit ($y=0$), 
and that on the BEC side ($y\gg 1)$ it rapidly converges to one. 

\section{Three-component ultracold fermions in two dimensions} 

A two-dimensional Fermi gas can be obtained by imposing a very strong 
confinement along one of the three spatial directions. In practice,  
the potential energy $E_p$ of this strong external confinement 
must be much larger than the total chemical potential $\mu_{3D}$ 
of the fermionic system: $\mu_{3D}\ll 2 E_P$ \cite{io-me}. 
Contrary to the three-dimensional case, 
in two dimensions quite generally 
a bound-state energy $\epsilon_B$ exists for any value 
of the interaction strength $g$ between atoms \cite{marini}. 
For the contact potential the bound-state equation is 
\beq 
- {1 \over g} = 
{1 \over V} \sum_{\bf k} \frac{1}
{{\hbar^2k^2\over 2m} + \epsilon_B} \, ,
\eeq 
and then subtracting this equation from 
the gap equation (\ref{bcsGap}) 
one obtains the two-dimensional regularized gap equation \cite{marini} 
\beq 
\sum_{\bf k} \left( 
\frac{1}{ {\hbar^2k^2\over 2m} + \epsilon_B} 
- {1\over 2 E_k} \right) = 0 \; . 
\label{bcs2}  
\eeq 
Note that, for a 2D inter-atomic potential described 
by a 2D circularly symmetric well of radius $R_0$ and 
depth $U_0$, the bound-state energy $\epsilon_B$ is given 
by $\epsilon_B \simeq \hbar^2/(2mR_0^2) \exp{(-2\hbar^2/(mU_0R_0^2))}$ 
with $U_0 R_0^2 \to 0$ \cite{landau}.  

In the two-dimensional 
continuum limit $\sum_{\bf k} \to V/(2\pi)^2 \int d^2{\bf k}
\to V/(2\pi) \int k dk$, the Eq. (\ref{bcs2}) gives 
\beq 
\epsilon_B = \Delta 
\left( \sqrt{1+{\mu^2\over \Delta^2}}- {\mu\over \Delta} \right) \; .  
\label{gap-2d}
\eeq
Instead, the number equation (\ref{bcs1}) with (\ref{bcs1a}) 
and (\ref{bcs1b}) gives the total number density as 
\beq 
n = {N\over V} = n_R + n_G + n_B \; ,
\eeq 
where $V$ is a two-dimensional volume (i.e. an area), the red and green  
densities are 
\beq 
n_R = n_G = {1\over 2} \big({m \over 2\pi\hbar^2}\big) \Delta 
\left( {\mu\over \Delta} + \sqrt{1+{\mu^2\over \Delta^2}} \right) \; ,  
\label{numb} 
\eeq
while the blue density reads 
\beq 
n_B = \big({m \over 2\pi\hbar^2}\big) \mu \ \Theta(\mu) \; . 
\eeq

\begin{figure}
\centerline{\epsfig{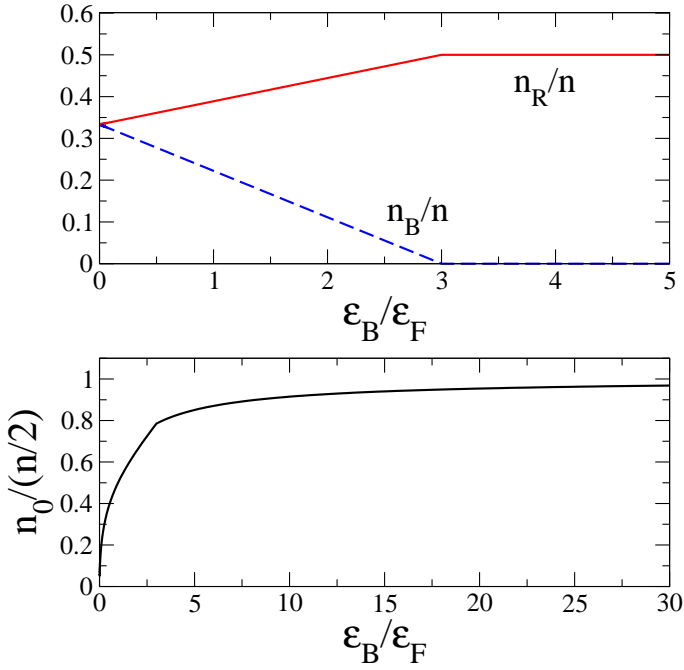}}
\small 
\caption{(Color online). Three-component ultracold 
fermions in two-dimensions. 
Upper panel: fraction of red fermions $n_R/n$ (solid line) 
and fraction of blue fermions $n_B/n$ (dashed line) 
as a function of scaled bound-state energy $\epsilon_B/\epsilon_F$. 
Lower panel: condensed fraction of red-green particles $n_0/n$ 
as a function of scaled bound-state energy $\epsilon_B/\epsilon_F$.} 
\label{fig2}
\end{figure} 

Finally, the condensate density of red-green pairs is given by 
\beq 
n_0 = {1\over 4} \big({m\over 2 \pi \hbar^2}\big) \Delta 
\left( {\pi\over 2} + \arctan{({\mu\over \Delta})} \right) \; . 
\label{cond}
\eeq 
Also in this two-dimensional case all the relevant quantities can be expressed 
in terms of the ratio $x_0 = {\mu/\Delta}$. In particular, 
the fraction of red fermions, which is equal to 
the fraction of green fermions, is given by 
\beq 
{n_R\over n} = {n_G\over n} = 
{x_0+\sqrt{1+x_0^2}\over 2 [x_0 + \sqrt{1+x_0^2}+x_0 \ \Theta(x_0) ] }  \; , 
\eeq 
the fraction of blue fermions is 
\beq 
{n_B\over n} = 1 - 2 {n_R\over n} \; , 
\eeq
and the condensate fraction is 
\beq 
{n_0\over n} = { {\pi\over 2} + \arctan{(x_0)} \over 
4[ x_0 + \sqrt{1+x_0^2} + x_0 \ \Theta(x_0) ] }  \; .   
\eeq 
In two-dimensions the Fermi energy $\epsilon_F=\hbar^2k_F^2/(2m)$ 
of a non-interacting Fermi gas is given by 
$\epsilon_F=\pi \hbar^2 n/m$ with $k_F=(4\pi N/(3V))^{1/2}=
(4\pi n/3)^{1/2}$ the Fermi wave number. 
It is convenient to express 
the bound-state energy $\epsilon_B$ in terms of the Fermi energy 
$\epsilon_F$. In this way we find 
\beq 
{\epsilon_B\over \epsilon_F} = 
3 { 
\sqrt{1+ x_0^2} - x_0   
\over 
x_0+ \sqrt{1+ x_0^2} + x_0\ \Theta(x_0) } \; ,  
\label{epsilonB} 
\eeq
We can now use these parametric formulas of $x_0$ to plot the fractions 
as a function of the scaled bound-state energy $\epsilon_B/\epsilon_F$. 

In the upper panel of Fig. \ref{fig2} we plot the fraction 
of red fermions $n_R/n$ (solid line) and the 
fraction of blue fermions $n_B/n$ (dashed line) 
as a function of scaled bound-state energy $\epsilon_B/\epsilon_F$.
The behavior of $n_G/n$ is not shown because it is exactly 
the same of $n_R/n$. The figure shows that 
in the deep BCS regime ($\epsilon_B/\epsilon_F \ll 1$) 
the system has $n_R/n=n_G/n=N_B/n=1/3$. 
By increasing $\epsilon_B/\epsilon_F$ 
the fraction of red and green fermions increases while 
the fraction of blue fermions decreases. At $\epsilon_B/\epsilon_F
= 3$, where $x_0=0$, the fraction of blue fermions becomes zero. For larger 
values of $\epsilon_B/\epsilon_F$ there are only 
the paired red and green particles. 
This behavior is quite similar to the one of the 
three-dimensional case; the main difference is due to the fact 
that here the curves are linear. 

In the lower panel of Fig.\ref{fig2} 
it is shown the condensate fraction $n_0/(n/2)$ of 
red-green pairs. 
In the weakly-bound BCS regime ($\epsilon_B/\epsilon_F \ll 1$) 
the condensed fraction $n_0/n$ goes to zero, while in the 
strongly-bound BEC regime ($\epsilon_B/\epsilon_F \gg 1$) 
the condensed fraction $n_0/n$ goes to $1/2$, i.e. all the 
red-green Fermi pairs belong to the Bose-Einstein condensate. 
Notice that the condensate fraction is zero 
when the bound-state energy $\epsilon_B$ is zero. 
For small values of $\epsilon_B/\epsilon_F$ the condensed 
fraction has a very fast grow but then it reaches the 
asymptotic value $1/2$ very slowly. 

\section{Inclusion of harmonic confinement} 

It is interesting to study the effect of a harmonic potential 
\beq 
U(r)= {1\over 2} m \omega^2 r^2 \;  
\eeq
on the properties of the three-component ultracold gas in the BCS-BEC 
crossover. For semplicity we investigate the two-dimensional case, 
which gives rise to elegant formulas also in this non-uniform configurtion. 
In fact, by using the local density approximation, namely 
the substitution 
\beq 
\mu \to \mu(r) = {\bar \mu} - U(r) \; , 
\eeq
the gap equation (\ref{gap-2d}) gives the space-dependent gap parameter as 
\beq 
\Delta(r) = \Delta_0 \ (1 - {r^2\over r_0^2}) \ 
\Theta(1-{r^2\over r_0^2}) 
\; , 
\eeq
where $\Delta_0 =\sqrt{\epsilon_B^2+2\epsilon_B \bar{\mu}}$ 
and $r_0=\Delta_0/\sqrt{\epsilon_B m \omega}$. Here $\bar{\mu}$ 
is the chemical potential of the non-uniform system. 

\begin{figure}
\centerline{\epsfig{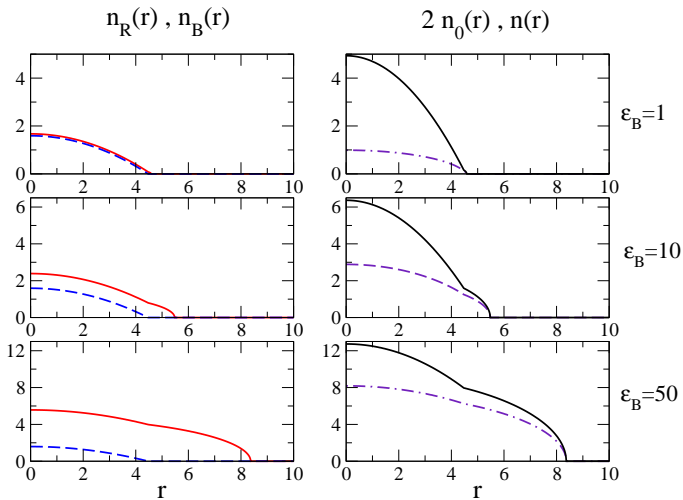}}
\small 
\caption{(Color online). Three-component ultracold fermions 
in two-dimensions under harmonic confinement $U(r)=r^2/2$. 
Left panels: density profile $n_R(r)$ of red fermions (solid lines) 
and density profile $n_B(r)$ of blue fermions (dashed lines). 
Right panels: total density profile $n(r)=2n_R(r)+n_B(r)$ 
(solid lines) and condensate density profile 
$2n_0(r)$ (dot-dashed lines). Note that the density $n_G(r)$
of green fermions is not shown because it is equal to $n_R(r)$. 
Data obtained with $\bar{\mu}=10$ 
and three values of the bound-state energy $\epsilon_B$: 
$\epsilon_B=1$ (upper panels), $\epsilon_B=10$ (middle panels), 
and $\epsilon_B=50$ (lower panels). 
We use scaled variables where lengths are  
in units of $a_H=\sqrt{\hbar/(m\omega)}$, with $\omega$ the 
frequency of harmonic confinement, and densities in units of $a_H^{-2}$.} 
\label{fig3}
\end{figure} 

In the same way the density profiles of red, green and blue 
fermions read 
\beq
n_R(r) = n_G(r) = {1\over 2} \big({m \over 2\pi\hbar^2}\big) \Delta(r)  
\left( {\mu(r)\over \Delta(r)} + \sqrt{1+{\mu(r)^2\over \Delta(r)^2}} 
\right) \; ,  
\eeq
\beq
n_B(r) = \big({m \over 2\pi\hbar^2}\big) \mu(r) \ \Theta(\mu(r)) \; . 
\eeq
The density profile of condensed red-geen pairs is instead given by 
\beq 
n_0(r) = {1\over 4} \big({m\over 2 \pi \hbar^2}\big) \Delta(r)  
\left( {\pi\over 2} + \arctan{({\mu(r)\over \Delta(r)})} \right) \; . 
\eeq

In Fig. \ref{fig3} we plot results obtained by using scaled variables: 
energy in units of $\hbar\omega$ and length in units of 
$a_H=\sqrt{\hbar/(m\omega)}$. We work at fixed chemical potential 
$\bar{\mu}=10$ and increase the bound-state energy $\epsilon_B$. 
In the upper panels of Fig. \ref{fig3} we set $\epsilon_B=1$, 
in the middle panels $\epsilon_B=10$ and in the lower ones 
$\epsilon_B=50$. 
In the left panels of Fig. \ref{fig3} we show density profiles 
of red particles (solid lines) and blue particles (dashed lines). 
In the right panels we show instead the total density profile 
$n(r)=n_R(r)+n_G(r)+n_B(r)$ (solid line) and the density profile 
$2n_0(r)$ of condensed particles (dot-dashed lines). 
As expected, by increasing $\epsilon_B$ 
the critical radius $r_0=\sqrt{\epsilon_B+2\bar{\mu}}$ 
(written in scaled units), at which the gap order parameter 
$\Delta(r)$ and the condensate density $n_0(r)$ 
go to zero, becomes much larger than the critical radius 
$r_T=\sqrt{2\bar{\mu}}$, at which the density $n_B(r)$ of 
blue fermions goes to zero. Moreover, by increasing $\epsilon_B$ 
the cloud of blue particles becomes quite small while the 
cloud of condensed particles approaches the total one. 

\section{Conclusions}

We have investigated the condensate fraction and the population 
imbalance of a uniform three-component ultracold fermions by increasing 
the U(3) invariant attractive interaction. We have considered 
the superfluid system both in the three-dimensional case and in the 
two-dimensional one. For the uniform system 
we have obtained explicit formulas 
and plots for number densities, condensate density and 
population imbalance in the full BCS-BEC crossover. 
For the system under harmonic confinement we have analyzed 
the density profiles of the three components and the density 
profile of the condensed pairs by varying the 
interaction strength. In our calculations we have used the 
mean-field extended BCS theory. Monte Carlo simulations 
have shown that, at zero-temperature, beyond-mean-field effects 
are negligible in the BCS side of the BCS-BEC crossover while they become 
relevant in the BEC side \cite{astrakharchik,bertaina}. 
Nevertheless, in the deep BEC side the condensate fraction goes in any case 
to one and the main difference in its determination is around 
the unitarity limit, where the relative difference between mean-field 
and Monte Carlo condensate fraction is about 20\% \cite{astrakharchik}. 
We think that our results can be of interest for next future experiments 
with degenerate gases made of alkali-metal or alkaline-earth atoms. 
As stressed in the introduction, SU(N) invariant interactions 
can be experimentaly obtained by using these atomic 
species \cite{selim,hara,ana}. The problem of unequal couplings, 
and also that of a fixed number of atoms for each component, 
is clearly of big interest too, and its analysis can be 
afforded by including more than one order parameter 
(see for instance \cite{torma}). 

The author thanks Flavio Toigo for useful discussions and 
Nicola Manini for computational help.

\end{document}